\documentclass{aa}
\begin{document}
\title{Instabilities, turbulence, and mixing in the ocean of accreting 
neutron stars}
\author{V.~Urpin}
\offprints{V.~Urpin}
\institute{Departament de Fisica Aplicada, Universitat d'Alacant,
           Ap. Correus 99, 03080 Alacant, Spain \\
           A.F.Ioffe Institute for Physics and Technology and
           Isaak Newton Institute of Chili, Branch in St. Petersburg,
           194021 St Petersburg, Russia}
\date{today}
\abstract{
We consider the stability properties of the ocean of accreting
magnetic neutron stars. It turns out that the ocean is always
unstable due to the combined influence of the temperature and
chemical composition gradients along the surface and of the Hall
effect. Both the oscillatory and non-oscillatory
modes can be unstable in accreting stars. The oscillatory instability 
grows on a short timescale $\sim 0.1-10$ s depending on the 
lengthscale of a surface inhomogeneity and the wavelength of
perturbations. The instability of non-oscillatory modes is typically
much slower and can develop on a timescale of hours or days. 
Instability generates a weak turbulence that can be responsible for 
mixing between the surface and deep ocean layers and for spreading 
the accreted material over the stellar surface. Spectral 
features of heavy elements can be detected in the atmospheres of 
accreting stars due to mixing, and these features should be 
different in neutron stars with both stable and unstable burning. 
Motions caused by instability can also be the reason for slow 
variations in the luminosity.
}

\titlerunning{Instabilities in the ocean of accreting neutron stars}

\maketitle

\section{Introduction}

The surface layers of accreting neutron stars are in a melted state 
because of the heat production due to nuclear burning of the accreted 
material. The way a neutron star burns the accreted hydrogen and 
helium to heavier elements is sensitive to the accretion rate (e.g., 
Bildsten 1998). This nuclear burning is very to be stable for 
relatively high accretion rate, $\dot{M} \geq 2.6 \times 10^{-8} 
M_{\odot}$ yr$^{-1}$. Burning at such accretion rates is not hot 
enough to produce iron group or heavier elements; thus the surface 
layers of neutron stars consist mainly of CNO group elements (Brown 
\& Bildsten 1998). Fusion of this fuel to heavier elements does not 
occur at the density $\rho \leq 10^{9}$ g cm$^{-3}$. If the neutron 
star is accreting pure He then the burning at $\rho > 10^{9}$ g 
cm$^{-3}$ produces iron group elements, but if the star accretes a 
mixture of H and He then the products are much havier (Schatz et al 
1999). The low ionic charge and a relatively high temperature ($\geq 
10^{8}$K) caused by the nuclear burning delay cristallization until 
very high densities are reached. Therefore neutron stars accreting 
at a high accretion rate are covered with massive oceans. Neutron 
stars accreting at lower rates, $\dot{M} < 2.6 \times 10^{-8} 
M_{\odot}$yr$^{-1}$, burn accreted H and He directly onto iron group 
elements (Lewin, van Paradijs \& Taam 1992, Schatz et al. 2001, 
Woosley et al. 2004), while the Coulomb interaction between heavy 
ions leads to crystallization at much lower densities and to a much 
shallow ocean.

MHD processes in the ocean are of general interest to studies of  
magnetic and thermal evolution, thermonuclear burning, neutron star
seismology, and properties of quasi-periodic oscillations. For instance, 
in weakly magnetized accreting neutron stars, the ocean of light 
elements supports shallow water waves, which can be the reason for low 
frequency ($\sim 5-7$ Hz) quasi-periodic oscillations in the brightest 
X--ray sources (Bildsten \& Cutler 1995, Bildsten, Ushomirski \& Cutler 
1996). Under certain conditions, large-scale flows in the ocean can 
influence the magnetic evolution of accreting neutron stars 
(Bisnovatyi-Kogan \& Komberg 1974, Rai Choudhuri \& Konar 2002). MHD 
phenomena can play a particular role during thermonuclear X-ray bursts. 
Thus, if an unstable nuclear burning is not spherically symmetric, 
then burning fronts can spread around the star by igniting the 
accumulated cold fuel ahead of the moving front (Fryxell \& Woosley 
1982, Bildsten 1993, Spitkovsky, Levin \& Ushomirsky 2001). The 
propagation speed can be considerably increased when convection and 
enhanced mixing are occurring at the burning front since convective 
motions are very efficient at transporting heat in the ocean. Note 
that effects caused by rotation should be important during X-ray 
bursts since most bursters are rapidly rotating. Most likely, an 
expanding ocean rotates differentially, and differential rotation can 
be the reason for MHD instabilities (Cumming \& Bildsten 2000, Menou
2004). 

Mixing in stars is often atributed to hydrodynamic instabilities, 
such as convection. Standard thermal convection arises if the 
temperature gradient is superadiabatic and occurs only in the 
atmosphere of hot NSs with a weak magnetic field (Miralles, Urpin \& 
Van Riper 1997). Convection, however, is not the only instability 
that occurs in the ocean. Stability properties of accreting neutron
stars can be complicated because of the presence of a strong magnetic 
field, $B \sim 10^{12}-10^{13}$ G, and of the thermal and compositional
gradients that generally are not parallel to gravity $\vec{g}$. A 
possibile occurrence of the Kruskal-Schwarzschild instability in polar 
caps of accreting neutron stars has been considered by Litwin, Brown \& 
Rosner (2001), who found that ballooning modes with the displacement 
perpendicular to the magnetic field can be unstable if the overpressure 
at the bottom of the neutron star ocean exceeds magnetic pressure by 
a factor $\sim 8 a/h$, where $a$ and $h$ are the horizontal lengthscale 
of the cap and the density scale hight, respectively. The crucial point 
of the instability is line-tying to the neutron star crust and, 
therefore, unstable modes are localized to within a density scale 
height of the ocean bottom. The ballooning instability can arise on a 
timescale $\sim 30$ hr if $\dot{M} \sim 10^{-10} M_{\odot}$/yr and, most
likely does not affect the nuclear burning but can degrade the 
confinement to prevent accumulation of mass above the instability 
threshold.

One more instability that can manifest itself in accreting neutron 
stars is the well-known Parker instability (Parker 1966). The downward 
flow of accreted matter may generate strong non-uniformity in the 
magnetic field and, hence, the electric current, which in turn can 
lead to Parker-type instability (Cumming, Zweibel \& Bildsten 2001). 
Note that this instability arises only if both the magnetic field and 
its non-uniformities are sufficiently strong and if the field decreases 
with height. Employing the effect of thermal diffusivity, Cumming, 
Zweibel \& Bildsten (2001) estimated that the critical field above which 
the instability occurs is $\sim 10^{12}$ G at the top of the crust. 
The Parker-type instability can be of particular importance if
the magnetic field in the surface layers has a complex geometry, 
for example, such as one considered by Urpin \& Gil (2004).

The material accreted from a companion star carries angular momentum. 
The inflow of angular momentum promotes differential rotation in the 
ocean of neutron stars and can cause rotation induced instabilities. 
These instabilities occur when the angular velocity gradient is so 
steep that the destabilizing effect of shear overwhelms the stabilizing 
effect of buoyancy (see, e.g., Zahn 1983). In the context of neutron 
star oceans, the instabilities caused by differential rotation have 
been considered by Fujimoto (1988, 1993). Turbulent motions generated 
by the instabilities transport the angular momentum in the ocean 
effectively, but the elemental mixing is much less efficient because 
it is hampered by a stable stratification. Note that a stable 
stratification can suppress some rotation induced instabilities since 
the Brunt-V\"{a}is\"{a}l\"{a} frequency is typically much larger than 
the angular velocity in accreting neutron stars (Cumming \& Bildsten 
2000). If differential rotation departs from the cylindrical symmetry,
then the ocean can be subject to baroclinic instability (Cumming \& 
Bildsten 2000). A stable stratification generally suppresses the modes 
with horizontal lengthscales that are much longer than a pressure scale 
height, but short-wavelength can be unstable, and turbulent transport
caused by the baroclinic instability can be significant during type 1 
X-ray bursts. Note also that some mixing in accreting neutron stars can 
be provided by tidal waves (Lou 2001). It is unlikely, however, that 
this mechanism is efficient since tidal waves in the neutron star 
ocean shoud have a small amplitude because of high gravity.      

In the present paper, we consider the instability caused by horizontal
inhomogeneities of the temperature and chemical composition in the ocean 
of accreting magnetic neutron stars. The thermal and compositional state 
of the ocean departs substantially from the spherical symmetry for such 
stars. The departure can be caused, for example, by the accreted 
hydrogen and helium focusing onto the magnetic poles. It is usually 
assumed that the accretion flow interacts with the magnetosphere at 
the so-called Alfv\'en radius, $R_{A}$, which is determined by the 
balance of the magnetic and inertial forces. For the standard neutron 
star with a dipole magnetic field and with mass $M=1.4 M_{\odot}$ and 
radius $R \approx 10^{6}$cm, the Alfv\'en radius is given by
\begin{equation}
R_{A} \sim 2 \times 10^{9} B_{13}^{4/7} \dot{M}_{-10}^{-2/7} \;
{\rm cm},
\end{equation}
where $B_{13} = B/10^{13}$G and $\dot{M}_{-10}=\dot{M}/(10^{-10} 
M_{\odot} {\mathrm yr}^{-1})$. For a weakly magnetized star accreting 
at a high accretion rate, the Alfv\'en radius becomes comparable to the 
stellar radius, and the magnetic field cannot funnel the accreted 
material onto the magnetic poles. Therefore, the distribution of the 
accreted material over the stellar surface is more or less smoothed in 
such stars. In contrast, a strongly magnetized neutron star accretes 
the material predominantly onto the magnetic poles, and distribution 
of the chemical composition and temperature is substantially 
non-spherical in the ocean of magnetized stars (Brown \& Bildsten 1998). 

It has been shown by Urpin (2004) that the combined influence of the 
Hall effect and a thermal gradient perpendicular to $\vec{g}$ can be 
responsible for instability in the ocean of isolated neutron stars. In 
accreting neutron stars, the presence of a compositional gradient that 
generally is also non-parallel to $\vec{g}$ can lead to similar 
instability. This sort of instability is qualitatively different 
from instabilities already considered in accreting neutron stars 
because it occurs only in the presence of the Hall effect. 
The criterion of instability can be satisfied in many accreting 
neutron stars if the compositional gradient is not parallel to gravity. 
 
This paper is organized as follows. In Section 2, we consider the main
equations governing hydrodynamic motions in the ocean of accreting
magnetic neutron stars and derive the dispersion equation. Section 3 
deals with the criterion and growth time of instability. In Section 4, 
we discuss mixing processes associated to compositional instability 
in the ocean. 

\section{The basic equations}

We consider the linear stability of the ocean on an accreting neutron 
star assuming that both the temperature and chemical composition can
generally depart from spherical symmetry in the unperturbed state.
We do not specify the mechanisms responsible for these departures
but they can originate, for example, in the magnetic field 
$\vec{B}$ that funnels the accretion flow onto the poles and 
provides anisotropy to the heat transport. As a result, gradients of 
the temperature and chemical composition are not parallel to gravity 
$\vec{g}$, which is approximately radial in the ocean. 

The depth of the ocean depends crucially on the temperature and 
chemical composition and varies over a wide range. Crystallization 
of a Coulomb liquid occurs when the ion coupling parameter $\Gamma = 
Z^{2} e^{2}/(a k_{B} T)$ reaches the critical value $\Gamma = \Gamma_{m} 
\approx 170-180$ (Slattery, Doolen \& De Witt 1980, Potekhin \&
Chabrier 2000); $a= (3/4 \pi n_{i})^{1/3}$ is the mean inter-ion 
distance, $n_{i}$ and $Z$ are the number density and the charge number 
of ions, respectively; $k_{B}$ is the Boltzmann constant. Then, the 
crystallization temperature is
\begin{equation}
T_{m} = \frac{Z^{2} e^{2}}{a k_{B} \Gamma_{m}} \approx
1.3 \times 10^{5} Z^{5/3} x^{1/3}
\left( \frac{170}{\Gamma_{m}} \right) \;\; K,
\end{equation}
where $x= Z \rho /10^{6}A$ g/cm$^{3}$, and $A$ is the atomic
number of ions. In neutron stars with temperature $T \geq 10^8$  
which is more or less typical for stars accreting with the rate $\sim
10^{-9} M_{\odot}$ yr$^{-1}$, crystallization occurs at the density 
$\rho \sim 3 \times 10^{10}$ g/cm$^3$ if $Z=8$.  

The equations governing velocity, magnetic field, and thermal balance 
are (see, e.g., Landau \& Lifshitz 1959) 
\begin{equation}
\dot{\vec{v}} + (\vec{v} \cdot \nabla) \vec{v} = - \frac{\nabla
p}{\rho} + \vec{g} + \frac{1}{4 \pi \rho} (\nabla \times \vec{B})
\times \vec{B}, 
\end{equation}
\begin{equation}
\dot{\rho} + \nabla \cdot (\rho \vec{v}) = 0, 
\end{equation}
\begin{equation}
\dot{\vec{B}} - \nabla \times (\vec{v} \times \vec{B}) = -
\nabla \times [\hat{\eta} \cdot ( \nabla \times 
\vec{B})], 
\end{equation}
\begin{equation}
\nabla \cdot \vec{B} = 0, 
\end{equation}
\begin{equation}
\dot{T} - \frac{\beta}{\rho c_{p}} \dot{p} + \vec{v} \cdot 
(\Delta \nabla T) = \frac{1}{\rho c_{p}} \nabla \cdot (\hat{\kappa} 
\cdot \nabla T),
\end{equation}
where $\vec{v}$ is the fluid velocity; $\Delta \nabla T = \nabla T - 
\nabla_{ad}T$ is the difference between the real and adiabatic 
temperature gradients, $\nabla_{ad} T = \beta \nabla p/ \rho c_{p}$; 
$\beta = - (\partial \ln \rho/\partial \ln T)_{p}$ is the thermal 
expansion coefficient and $c_{p}$ the specific heat at constant 
pressure, $p$; and finally $\hat{\eta} = c^{2} \hat{R}/4 \pi$, 
$\hat{R}$ and $\hat{\kappa}$ are tensors of the electrical 
resistivity and thermal conductivity. Following some tensor 
operations on Eqs.~(5) and (7), we have
$$
\hat{\kappa} \cdot \nabla T = \kappa^{(0)} \vec{b} (\vec{b}
\cdot \nabla T) +
\kappa^{(B)} [\nabla T - \vec{b} (\vec{b} \cdot \nabla T)] + 
\kappa^{(\wedge)} \vec{b} \times \nabla T,
$$
where $\kappa^{(0)}$ and $\kappa^{(B)}$ are the tensor 
components along and across the magnetic field, respectively, 
$\kappa^{(\wedge)}$ is the so called Hall component, and $\vec{b} 
= \vec{B}/B$.  An analogous expression can be written for $\hat{\eta}$. 
Note that for the magnetic diffusivity we have $\eta^{(0)}= \eta^{(B)} 
= \eta^{(\wedge)}/ \alpha$ where $\alpha$ is the Hall parameter,
\begin{equation}
\alpha = \Omega_{Be} \tau \approx \frac{9.9 \times 10^{3} B_{13}}{Z
\Lambda (1+x^{2/3})}, 
\end{equation} 
$\Omega_{Be}$ and $\tau$ are the electron gyrofrequency and relaxation 
time, respectively; and $\Lambda$ is the Coulomb logarithm.

We neglect the viscous term in the momentum equation (3) since 
viscosity plays a less important role among kinetic processes in 
the ocean. Usually, both electrons and ions contribute to the shear 
viscosity. The ion shear viscosity dominates at a relatively low 
density but is typically much smaller than magnetic diffusivity 
$\eta$ and thermal diffusivity, $\chi= \kappa/ \rho c_{p}$. 
Electron viscosity is greater in deep layers of the ocean 
(Itoh, Kohyama \& Takeuchi 1987). The  ratio of electron 
viscosity $\nu_e$ and magnetic diffusivity along the magnetic 
field is given by
\begin{equation}
\frac{\nu_e}{\eta^{(0)}} = \frac{2.2 \times 10^{2}}{AZ \Lambda} \;
\frac{x^{5/3}}{(1 + x^{2/3})^{2}} ,
\end{equation} 
where $\Lambda$ is the Coulomb logarithm. The kinematic viscosity 
becomes comparable to the magnetic one only at a high density 
$\geq 10^{9}$ g/cm$^{3}$ if the ocean consists of light elements. 
Therefore, neglecting viscosity in Eq.~(3) seems to be 
qualitatively justified. 

We assume that the ocean is in hydrostatic equilibrium in the 
unperturbed state,
\begin{equation}
\frac{\nabla p}{\rho} = \vec{G} = \vec{g} + \frac{1}{4 \pi \rho} 
(\nabla \times \vec{B}) \times \vec{B}. 
\end{equation}
Taking the curl of this equation, we have
\begin{equation}
\nabla \rho \times \vec{G} = - \frac{1}{4 \pi} \nabla \times 
[(\nabla \times \vec{B}) \times \vec{B}] .
\end{equation} 
The density and, hence, temperature and compositional gradients
have components perpendicular to $\vec{G}$ if the magnetic
field is not force-free in the ocean.

Consider only relatively fast hydrodynamic processes varying on 
a timescale shorter than the characteristic timescale of nuclear 
reactions. The time to burn all hydrogen into CNO-group elements
is, for example, on the order of 20 hours at high accretion rates
(see, e.g., Cumming \& Bildsten 2000). Under this assumption, the number 
density of a species $j$ satisfies the continuity equation
\begin{equation}
\frac{\partial n_j}{\partial t} + \nabla \cdot (n_j \vec{v}) = 0.
\end{equation} 
For our purposes, it will be convenient to characterize the composition
by a mass fraction of species $X_j = \rho_j/\rho$, where $\rho_j$ is 
the density of species $j$. Note that if the number of species is $N$ 
then only $N-1$ fractions $X_j$ are linearly independent, since $X_j$ 
satisfy the condition $\sum_{N} X_j =1$; summation is over all species 
of ions. If the chemical composition is frozen in a moving fluid parcel, 
then the equations governing $X_j$ read
\begin{equation}
\frac{\partial X_j}{\partial t} + \nabla \cdot ( X_j \vec{v}) = 0.
\end{equation}

The equations for small perturbations can be obtained by linearization 
of Eqs.~(3)--(7), and (13). We assume that in the unperturbed state, a 
steady temperature and composition gradients are maintained, and there 
are no motions. In what follows, small perturbations will be marked by 
the subscript 1, but subscripts will be omitted for unperturbed 
quantities. We consider short-wavelength perturbations with spatial 
and temporal dependence $\exp (\gamma t - i \vec{k} \cdot \vec{r})$ 
where $\vec{k}$ is the wave-vector. Linearization of Eqs.~(3)--(7), 
(13) yields
\begin{equation}
\gamma \vec{v}_{1} =  i \vec{k} \frac{p_{1}}{\rho} + \vec{G} 
\frac{\rho_{1}}{\rho} - \frac{i}{4 \pi \rho} (\vec{k} \times \vec{B}_{1})
\times \vec{B},  
\end{equation}
\begin{equation}
\gamma \rho_{1} + \vec{k} \cdot \vec{v}_{1} = 0, 
\end{equation}
\begin{equation}
(\gamma + \omega_{\eta}) \vec{B}_{1} = - i (\vec{k} \cdot \vec{B}) 
\vec{v}_{1} - \eta_{\wedge} (\vec{k} \cdot \vec{b}) \vec{k} \times 
\vec{B}_{1},
\end{equation}
\begin{equation}
\vec{k} \cdot \vec{B}_{1} = 0, 
\end{equation}
\begin{equation}
(\gamma + \omega_{\kappa}) T_{1} - \frac{\gamma \beta}{\rho c_{p}} p_{1}
= - \vec{v}_{1} \cdot (\Delta \nabla T),
\end{equation}
\begin{equation}
\gamma X_{1j} + \vec{v}_{1} \cdot \nabla X_{j} = 0,
\end{equation}
where the characteristic inverse timescales of the ohmic dissipation 
and thermal diffusion are given by $\omega_{\eta} = \eta^{(0)} k^2$ and 
$\omega_{\kappa}= [\kappa^{(B)} k^2 + (\kappa^{(0)} - \kappa^{(B)}) 
(\vec{k} \cdot \vec{b})^2] / \rho c_{p}$, respectively. 

Perturbations of the pressure in Eq.~(14) can be expressed in terms of 
perturbations of the density, temperature, and fractions of species 
$X_{1j}$, using the equation of state,
\begin{equation}
\frac{p_{1}}{p} = \left( \frac{\partial \ln p}{\partial \ln \rho} 
\right)_{T, Y}
\left( \frac{\rho_{1}}{\rho} + \beta \frac{T_{1}}{T} + \sum^{N-1} 
\delta_{j} 
X_{1j} \right),
\end{equation} 
where $\delta_{j}= - (\partial \ln \rho/ \partial X_{j})_{pT}$ are the 
coefficients of chemical expansion, and summation is over $N-1$ linearly 
independent species.

Consider Eqs.~(14)-(20) in the case $k c_{s} \gg \gamma$ that corresponds 
to the Boussinesq approximation for slowly varying modes. Then, the 
dispersion equation reads
\begin{equation}
\gamma^{6} + a_{5} \gamma^{5} + a_{4} \gamma^{4} + a_{3} \gamma^{3} 
+ a_{2} \gamma^{2} + a_{1} \gamma + a_{0} = 0 \;, 
\end{equation}
where 
\begin{eqnarray}
&& a_{5} = \omega_{\kappa} + 2 \omega_{\eta} , \nonumber \\
&& a_{4} = \omega_{\eta}^2 + \omega_{\wedge}^2 + 2 \omega_{\eta} \omega_{\kappa}
+ 2 \omega_{A}^2 - \omega_{0}^2 , \nonumber \\
&& a_{3} = \omega_{\kappa} ( \omega_{\eta}^2 + \omega_{\wedge}^2 +
2 \omega_{A}^2 - \omega_{X}^2) + 2 \omega_{\eta} (\omega_{A}^2 -
\omega_{0}^2) , \nonumber \\
&& a_{2} = 2 \omega_{\eta} \omega_{\kappa} (\omega_{A}^2
- \omega_{X}^2) - \omega_{0}^{2} (\omega_{\eta}^2 + \omega_{\wedge}^2 + \omega_{A}^2)
+ \omega_{A}^4 , \nonumber \\
&& a_{1} = \omega_{A}^2 [ \omega_{\kappa} \omega_{A}^2 - 
\omega_{\eta} (\omega_{0}^2 +
\omega_{H}^2 ] 
- \omega_{\kappa} \omega_{Y}^2
(\omega_{\eta}^2 + \omega_{\wedge}^2 + \omega_{A}^2) , \nonumber \\
&& a_{0} = - \omega_{\kappa} \omega_{\eta} \omega_{A}^2 (\omega_{X}^2
+ \omega_{X H}^2). \nonumber 
\end{eqnarray}
The characteristic frequencies in this equation are
\begin{eqnarray}
&& \omega_{\wedge} = \alpha \eta_{\parallel} k (\vec{k} \cdot \vec{b}) \;, \;\;
\omega_{A} = \frac{(\vec{k} \cdot \vec{B})}{\sqrt{4 \pi \rho}} \;, 
\;\; \omega_{X}^2 = \vec{D} \cdot \nabla X \;, 
\nonumber \\
&& \omega_{g}^2 = \frac{\beta}{T} \vec{D} \cdot \Delta \nabla T \;,\;\; 
\omega_{0}^{2} = \omega_{g}^{2} + \omega_{X}^{2} = \beta \vec{D}
\cdot \vec{C} \;,
\nonumber \\
&& \omega_{H}^{2} = \omega_{gH}^{2} + \omega_{XH}^{2} \;, \;
\omega_{g H}^2 = \frac{\alpha \beta (\vec{k} \cdot \vec{b})}{k^2 T} 
\Delta \nabla T \cdot (\vec{k} \times \vec{D}) \;, \nonumber \\
&& \omega_{X H}^2 = \frac{\alpha (\vec{k} \cdot \vec{b})}{k^{2}}
(\vec{k} \times \vec{D}) \cdot \nabla X \;,
\nonumber  
\end{eqnarray}
where vectors $\vec{D}$ and $\vec{C}$ are given by
$$
\vec{D}= \vec{G} - \frac{\vec{k}}{k^2} \; (\vec{k} \cdot \vec{G}) \;,
\;\; \vec{C} = \frac{\Delta \nabla T}{T} + \frac{\nabla X}{\beta} \;,
$$
and we denote 
$\nabla X = \sum^{N-1} \delta_{j} \nabla X_{j}$. In the limit 
$\omega_{X} \approx \omega_{X H} \approx 0$, Eq.~(21) yields 
the dispersion equation for a chemically homogeneous ocean derived 
by Urpin (2004).  

\section{The growth rate and criterion of instability}

Equation (21) describes six modes that exist in a chemically 
inhomogeneous ocean of magnetic neutron stars. Stability properties 
of the modes depend crucially on $\vec{k}$ and, under certain 
conditions, modes can be unstable. The dissipative frequences 
$\omega_{\eta}$, $\omega_{\wedge}$, and $\omega_{\kappa}$ are typically 
smaller than the dynamical frequences $\omega_{A}$, $\omega_{0}$, and 
$\omega_{H}$ except perturbations with a very short wavelength, 
$\lambda=2 \pi/k$. Therefore, the dispersion relations for some modes 
can be obtained by making use of the perturbation method and assuming 
that dissipative effects are small. We expand the growth rate as 
$\gamma = \gamma^{(0)}+ \gamma^{(1)} + \gamma^{(2)} + ...$ where 
$\gamma^{(0)}$, $\gamma^{(1)}$, and $\gamma^{(2)}$ are terms 
of the zeroth, first, and second order in dissipative frequencies, 
respectively. The corresponding expansion should be made for the 
coefficients of Eq.~(21) as well. In the zeroth order when 
dissipation is neglected, Eq.~(21) reduces to a quadratic 
equation,
\begin{equation}
\gamma^{(0) \; 4} + (2 \omega_{A}^{2} - \omega_{0}^{2}) \gamma^{(0) \; 2} 
+ \omega_{A}^{2} ( \omega_{A}^{2} - \omega_{0}^{2})= 0, 
\end{equation} 
This equation describes four modes with the frequencies
\begin{equation}
\gamma^{(0) \; 2}_{1,2} = - \omega_{A}^{2}, \;\;\;
\gamma^{(0) \; 2}_{3,4} = - \omega_{A}^{2} + \omega_{0}^{2}.
\end{equation}
The fifth and sixth roots of Eq.~(21) are vanishing at the zeroth 
order of approximation, $\gamma_{5}^{(0)}=\gamma_{6}^{(0)}= 0$. 
Usually, $\gamma^{(0) \; 2} < 0$ for the modes 1-4 because the 
inequality $\omega_{0}^{2} > \omega_{A}^{2}$ requires temperature 
and composition gradients that do not exist in neutron stars. Therefore, 
modes 1-4 are oscillatory, and their instability does not occur at 
the zeroth order.

For applicability of the perturbation procedure, the first 
corrections to $\gamma^{(0)}$ caused by dissipative effects 
have to satisfy the condition $|\gamma^{(1)}| \ll |\gamma^{(0)}| 
\sim \omega_{A}$, since $\omega_{A}$ is usually larger than other 
frequencies. The real parts of $\gamma^{(1)}$ for the modes 1-4 
have a particularly simple form if $|\omega_{0}|$ and $|\omega_{H}|$ 
satisfy the inequalities
\begin{equation}
|\omega_{0}^{2}| \gg \omega_{\eta} \omega_{\kappa}(1 + \alpha^{2})
\; ,
\;\;\; |\omega_{H}^{2}| \gg \omega_{\eta} \omega_{\kappa} (1 + 
\alpha^{2}).
\end{equation}
Under this condition,
the real parts of $\gamma^{(1)}$ are given by
\begin{eqnarray}
&& {\mathrm Re} \; \gamma^{(1)}_{1, 2} = \frac{\omega_{\eta} 
( \omega_{H}^{2} - \omega_{0}^{2})}{2 \omega_{0}^{2}}, \\
&& {\mathrm Re} \; \gamma^{(1)}_{3, 4} = - \frac{\omega_{\eta} 
\omega_{A}^{2} [ \omega_{H}^{2} + \omega_{0}^{2} 
(1 - \zeta)]}{2 \omega_{0}^{2} (\omega_{A}^{2} - \omega_{0}^{2})},
\end{eqnarray}
where $\zeta = \omega_{\kappa} \omega_{0}^{2}/ \omega_{\eta} 
\omega_{A}^{2}$. Note that if condition (24) is not fulfilled
and dissipative effects dominate the dynamical influence of
temperature and composition gradients along the surface, then  
modes 1-4 are stable.

The roots 5 and 6 of Eq.~(21) describe a couple of non-oscillatory 
modes and are linear in dissipative effects. To calculate these roots 
we should keep the terms in Eq.~(21) up to the second order in
dissipative frequencies assuming that $\gamma$ is linear at these
frequencies. Then, roots 5 and 6 are approximately
\begin{equation}
\gamma_{5,6}^{(1)} \approx - \frac{A_{1}}{2} \pm \sqrt{\frac{A_{1}^{2}}{4}
-A_{0}},
\end{equation} 
where
\begin{eqnarray}
&&A_{0} = - \frac{\omega_{\kappa} \omega_{\eta} (\omega_{X}^{2} + 
\omega_{X H}^{2})}{\omega_{A}^{2} - \omega_{0}^{2}},
\nonumber \\ 
&&A_{1} = \frac{\omega_{\kappa} (\omega_{A}^{2} - \omega_{X}^{2}) -
\omega_{\eta} (\omega_{0}^{2} + \omega_{H}^{2})}{\omega_{A}^{2} - 
\omega_{0}^{2}}.
\nonumber 
\end{eqnarray}
Modes 5 and 6 are secular and appear only due to dissipative 
effects. Note that contrary to modes 1-4, expression (27) for 
non-oscillatory modes applies even for small wavelengths which do 
not satisfy the inequality (24). The only condition for applying 
Eq.~(27) is $|\gamma^{(1)}_{5, 6}| \ll \omega_{A}$.

\subsection{Instability of the oscillatory modes}

The real parts of oscillatory modes are given by Eqs.~(25) and
(26). Note that the vertical component of $\nabla X$ can be rather 
large because the chemical composition varies substantially with the 
ocean depth. Therefore, the contribution of $\nabla X$ to 
$\omega_{0}^{2}$ can usually be comparable to (or even larger than)
that of $\Delta \nabla T$. As mentioned, usually $\omega_{A}^{2} > 
\omega_{0}^{2}$ (see Eq.~(23)). If $\omega_{0}^{2} > 0$ then 
instability of oscillatory modes occurs if 
\begin{equation}
\omega_{H}^{2} > \omega_{0}^{2} \;\; {\rm or} \;\;
\omega_{H}^{2} < \omega_{0}^{2} (\zeta-1).
\end{equation}  
In contrast, if stratification is convectively stable and 
$\omega_{0}^{2} < 0$ then these modes are unstable if  
\begin{equation}
\omega_{H}^{2} < \omega_{0}^{2} \;\; {\rm or} \;\; 
\omega_{H}^{2} > \omega_{0}^{2} (\zeta -1).
\end{equation} 

In the particular case when gradients of the temperature and chemical 
composition are parallel to $\vec{G}$ and $\omega_{g H}^{2}= 
\omega_{XH}^{2}= 0$, modes 1 and 2 are stable but modes 3 and 
4 can be unstable if $\zeta > 1$, or
\begin{equation}
\omega_{\kappa} \omega_{0}^2 > \omega_{\eta} \omega_{A}^2.
\end{equation}
This condition generalizes the well-known criterion of oscillatory 
convection (see, e.g., Chandrasekhar 1961) for the case of a chemically 
inhomogeneous fluid. In the ocean of neutron stars, parameter 
$\zeta$ can be estimated as
\begin{equation}
\zeta \sim 3 \times 10^{-3} \frac{T_{7} \lambda_{2}^{2}}{Z \Lambda^{2} 
B_{13}^{2} H_{3}}
\frac{x^{2} [1 + (1 + x^{2/3})^{1/2}]}{(1 + x^{2/3})^{2}},
\end{equation}
where $H = C_{\parallel}^{-1} = (H_{T}^{-1} + H_{X}^{-1})^{-1}$, 
$H_{T} \equiv |(\Delta \nabla T/ T)_{\parallel}^{-1}|$, and $H_{X} = 
\beta |(\nabla X)_{\parallel}|^{-1}$ are the lengthscales 
of the temperature and chemical composition along $\vec{G}$ (the 
subscript $\parallel$ denotes the component parallel to $\vec{G}$), 
$H_{3} = H/ 10^{3}$ cm; and $\lambda_{2} = \lambda/ 100$ cm. Usually 
$\zeta < 1$, and condition (30) is not satisfied except for very 
hot stars with a low magnetic field (Miralles, Urpin \& Van Riper 
1997). 

If $|\zeta| \ll 1$, then conditions (28) and (29) read
\begin{equation}
\mid \omega_{H}^2 \mid > \mid \omega_{0}^2 \mid.
\end{equation} 
This inequality can always be satisfied by a corresponding choice of 
$\vec{k}$ once $(\Delta \nabla T)_{\perp} \neq 0$ or $(\nabla X)_{\perp} 
\neq 0$, where $\perp$ denotes the component perpendicular to $\vec{G}$. 
Introducing local coordinates with the $z$-axis antiparallel to 
$\vec{G}$ and the $x$-axis aligned with $\vec{k}_{\perp}$, we have 
$\vec{G}=- G \vec{e}_{z}$, $\vec{k} = k_{x} \vec{e}_{x} + k_{z} 
\vec{e}_{z}$; $\vec{e}_x, \vec{e}_y$, and $\vec{e}_z$ are unit 
vectors. The expressions for $\omega_{0}^{2}$ and $\omega_{H}^{2}$
read in these coordinates
\begin{equation}
\omega_{0}^{2} = \frac{k_x}{k^{2}} \beta G (k_z C_x - k_x C_z) \;,
\;\; \omega_{H}^{2} = \frac{k_x}{k^{2}} \alpha \beta G (\vec{k} \cdot 
\vec{b}) C_y \;,
\end{equation} 
where $C_{x, y, z}$ are the corresponding components of $\vec{C}$.
Then, condition (32) yields 
\begin{equation}
\alpha \mid \vec{k} \cdot \vec{b} \mid \; \mid C_{y} 
\mid \; > \; \mid k_x C_{z} - k_z C_{x} \mid ,
\end{equation}
The instability occurs only if $\vec{k}$ has a non-vanishing component 
perpendicular to the plane $(\vec{G}, \vec{C})$. Perturbations are 
suppressed if $\vec{k}$ is parallel to this plane and, hence, $C_y = 0$. 

Equation (34) applies only if condition (24) is satisfied and
\begin{equation}
|k_x C_z - k_z C_x | \gg 
\frac{k^{2} \omega_{\eta} 
\omega_{\kappa}}{\beta G k_x} \; (1 + \alpha^{2}).
\end{equation}
If inequality (35) is satisfied then we have for unstable perturbations
\begin{equation}
\alpha | \vec{k} \cdot \vec{b} | C_y \gg \frac{k^{2} \omega_{\eta} 
\omega_{\kappa}}{\beta G k_x} \; (1 + \alpha^{2}).
\end{equation}
Note the important difference between Eqs.~(34) and (35). Condition 
(34) is linear in $\vec{k}$, so it can always be satisfied by proper 
choice of the direction of $\vec{k}$. On the contrary, inequality 
(35) is non-linear and yields restrictions on the wavelength of 
unstable perturbations rather than on their direction. 

Consider initially condition (34). This condition is obviously
fulfilled if the wavevector is close to the direction determined 
by
\begin{equation}
\frac{k_{x}}{k_{z}} = \frac{C_{x}}{C_{z}} 
\sim \frac{H}{L},
\end{equation}
where $L = C_{\perp}^{-1} \sim (L_{T}^{-1} + L_{X}^{-1})^{-1}$,
$L_{T} \equiv |(\Delta \nabla T/ T)_{\perp}^{-1}|$, and $L_{X} = 
\beta |(\nabla Y)_{\perp}|^{-1}$ are the temperature and composition
lengthscales perpendicular to $\vec{G}$. One can estimate the angle 
$\Delta \theta$ around this direction where inequality (34) is 
still satisfied. Restricting ourselves by linear terms in $\Delta 
\theta$, we have from Eq.~(34)
\begin{equation}
\Delta \theta < \alpha \frac{|C_{y}|}{|C_{z}|} \left| b_{z}
+ b_{x} \frac{C_{x}}{C_{z}} \right|.
\end{equation} 
Note that the components $C_x$ and $C_y$ are usually  
small compared to $C_z$, 
\begin{equation}
| C_z| \sim \frac{1}{H} \gg |C_{x}| \sim | C_{y}| \sim \frac{1}{L}. 
\end{equation}
Therefore, Eq.~(38) is approximately equivalent to
\begin{equation}
\Delta \theta < \alpha |b_{z}| \frac{|C_{y}|}{|C_{z}|}
\sim \alpha |b_{z}| \frac{H}{L}
\end{equation} 
except the region near the magnetic equator where $b_{z} \approx 0$.
This estimate is correct if $\alpha < L/H$, or 
\begin{equation}
B < 2 \times 10^{11} Z L_{5} H_{3}^{-1} (1 + x^{2/3}) \;\;\; {\rm G},
\end{equation}
where $L_{5} = L/10^{5}$ cm.
Note, that for all perturbations with $\vec{k}$ within angle 
$\Delta \theta$ we have $k_{x} < k_{z}$. If magnetization is very 
strong and $\alpha \geq L/H$ then inequality (34) can be satisfied 
even for perturbations with $k_{x} \sim k_{z}$. 

Condition (35) determines the wavelength of unstable perturbations. 
Estimating $C_x \sim 1/L$, we can transform Eq.~(35) into
\begin{equation}
\frac{1}{L} > \frac{k}{k_x} \; \frac{\omega_{\eta} 
\omega_{\kappa}}{\beta G} \; f(\alpha) \;, \;\;\;
f(\alpha) = 1 +\alpha^{2} \;
\end{equation} 
(we assume that $\vec{k}$ is not perpendicular to $\vec{B}$).
Using the definitions of $\omega_{\eta}$ and $\omega_{\kappa}$ and
taking account of Eq.~(37), this expression can be rewritten as 
the condition for the wavelength,
\begin{eqnarray}
\lambda > \lambda_{cr} =  2 \pi L^{1/4} \left( \frac{k}{k_x}  
\right)^{1/4}
\left( \frac{\eta_{0} \kappa_{0}}{\beta G} \right)^{1/4} 
f^{1/4}(\alpha)
\nonumber \\ 
\sim 0.6 \; f^{1/4}
\left( \frac{Z^{2} L_{5}^{2}}{H_{3} x^{1/3}} \right)^{1/4} 
[1 + (1+x^{2/3})^{1/2}]^{-1/4} {\rm cm};
\end{eqnarray}
we use the electron thermal conductivity by Urpin \& Yakovlev (1980). 
For applicability of a short wavelength approximation, $\lambda$
and $\lambda_{cr}$ should be smaller than the vertical
lengthscale in the ocean, $H \sim 10-20$m. This condition can 
in general be satisfied for a wide range of parameters except in
the case of a very strong magnetic field. In such field $\alpha \gg 1$, 
and we have $f \approx \alpha$. Then, critical wavelength 
$\lambda_{cr}$ becomes larger than $H$ if 
\begin{eqnarray}
B > B_{d} \sim  4 \times 10^{15} 
\left( \frac{H_{3}}{L_{5}^{2}} \right)^{1/2} 
x^{1/6}(1+x^{2/3}) 
\nonumber \\
\times [1+(1+x^{2/3})^{1/2}]^{1/2} 
{\rm G}.
\end{eqnarray} 
Critical field $B_{d}$ is much stronger than typical magnetic fields
of the majority of pulsars.

The growth rate of instability depends on the wavelength and can 
vary within a wide range. If $\omega_{A}^{2} \gg \omega_{0}^{2}$ 
and $\zeta < 1$, then Re$\gamma$ for unstable modes is given by
\begin{equation}
{\rm Re} \; \gamma \approx \pm \frac{\omega_{\eta}}{2 \omega_{0}^{2}}
(\omega_{H}^{2} \mp \omega_{0}^{2}) =
\pm \frac{\omega_{\eta}}{2} \left[ \frac{\alpha (\vec{k} \vec{b})
C_y}{k_x C_z - k_z C_x} \mp 1 \right] .
\end{equation} 
The growth rate is maximal if $\vec{k}$ lies in the plane perpendicular 
to the plane $(\vec{G}, \vec{C})$. Then $C_{x} =0$ and, estimating 
$C_y / C_z \sim H/L$, from Eq.~(45) we have  
\begin{equation}
{\rm Re} \gamma \! \approx \! \frac{\eta_{0} k^{2}}{2}  
\frac{\alpha (\vec{k} \vec{b}) C_{y}}{k_x C_{z}} 
\sim 
10^{-3} B_{13} \epsilon^{-1} x^{-1} \lambda_{2}^{-2} 
\frac{H_{3}}{L_{5}} \;{\rm s}^{-1} \!\!.
\end{equation}  
The growth rate increases rapidly with increasing $k$ and reaches 
its maximum at $\lambda \sim \lambda_{cr}$. According to Eq.~(37), 
the minimal value of the ratio $\epsilon = k_x/k$ is on the order of 
$H/L$. Substituting this estimate into Eq.~(46) and assuming, 
for example, that $\lambda_2 \sim 1$, we have
\begin{equation}
{\rm Re} \gamma \sim 0.1 \; B_{13} x^{-1} \;\;{\rm s}^{-1}.
\end{equation}
The growth time for such perturbations is $\sim 10$ s in the layer 
with $\rho \sim 10^{6}$ g/cm$^{3}$ if $B \sim 10^{13}$G. Perturbations 
with a shorter vertical wavelength can grow even faster. For example,
perturbations with $\lambda \sim \lambda_{cr} \sim 10$ cm grow on
a timescale of about 0.1 s.

\subsection{Instability of the non-oscillatory modes}

Root (27) describes a couple of non-oscillatory modes that 
can generally be unstable in the ocean. The instability conditions
of these modes are
\begin{equation}
A_{0} < 0 \;\;\; {\rm or} \;\;\; A_{1} < 0.
\end{equation}
If we assume $\omega_{A}^{2} \gg \omega_{0}^{2}$, which is typical
for those neutron stars with $B > 3 \times 10^{8} x^{1/2} \lambda_{2} 
H_{3}^{1/2} L_{5}^{-1}$ G, then conditions (48) read
\begin{equation}
\omega_{X}^{2} + \omega_{XH}^{2} > 0,
\end{equation} 
or
\begin{equation}
\omega_{\kappa} (\omega_{A}^{2} - \omega_{X}^{2}) - 
\omega_{\eta} ( \omega_{0}^{2} + \omega_{H}^{2}) < 0.
\end{equation}

Consider initially the condition (49). If the magnetic field is
weak or $\vec{G}$ and $\nabla X$ are almost parallel (or antiparallel), 
we have $|\omega_{X}^{2}| \gg |\omega_{XH}^{2}|$, and inequality (49) 
transforms into $\omega_{X}^{2} > 0$. This condition has exactly the 
same form as the standard criterion of the Ledoux convection. Note, 
however, the big difference because the standard Ledoux convection is 
a dynamical instability, whereas the instability considered in this 
subsection is secular. Using the local cartesian coordinates, we can 
represent condition $\omega_{X}^{2} > 0$ as
\begin{equation}
\nabla_{z} X \equiv \sum_{j=1}^{N-1} \delta_{j} \frac{d X_{j}}{dz} < 0.
\end{equation}
As an example, in the simplest case of a mixture of two species with the
atomic masses $m_{1}$ and $m_{2}$, Eq.~(52) reads $\delta_{1} 
dX_{1}/ dz < 0$. If $m_{2} > m_{1}$ then $\delta_{1} = - (\partial 
\rho/ \partial X_{1})_{pT}$ is positive, and the condition of instability 
is $dX_{1}/dz < 0$, or $d(n_{1}/n_{2})/dz < 0$ where $n_{1}$ and $n_{2}$
are the number density of species 1 and 2, respectively. Therefore,
the instability occurs if the relative number density of a light
species increases with depth. Most likely, however, the relative
concentration of heavy elements increases with depth and 
$\omega_{X}^{2} < 0$. Therefore, the instability most likely does not 
occur in the ocean if $\nabla X \parallel \vec{G}$.

If $\nabla X$ and $\vec{G}$ are not parallel, then the stability
properties change drastically. In the general case, inequality 
(49) yields
\begin{equation}
\vec{G} \cdot \nabla X - \frac{1}{k^{2}} (\vec{k} \cdot \vec{G})
(\vec{k} \cdot \nabla X) - \frac{\alpha (\vec{k} \cdot \vec{b})}{k^{2}}
\; \vec{G} \cdot (\vec{k} \times \nabla X) > 0.
\end{equation}
Again using the local cartesian coordinates, we have from this 
equation 
\begin{equation}
k_{x} k_{z} (\alpha b_{z} \nabla_{y} X 
+ \nabla_{x} X ) > \; k_x^{2} (\nabla_{z} X - \alpha b_{x} \nabla_{y} X),  
\end{equation}  
where $\nabla_{x, y, z} X$ are cartesian components of $\nabla X$.
If the composition varies over the surface, then for any distribution 
of chemicals, there exists a region in the $(k_{x}, k_{z})$-plane 
where the condition of instability (53) is satisfied. The 
non-oscillatory modes become unstable if even one of the components 
$\nabla_{x} X$ or $\nabla_{y} X$ is non-vanishing. If the vertical 
component of a chemical gradient is large and $\nabla_{z} X > \alpha 
b_{x} \nabla_{y} X$, then the instability occurs if
\begin{equation}
\left| \frac{k_{x}}{k_{z}} \right| < \left| \frac{\alpha b_{z} 
\nabla_{y} X
+ \nabla_{x} X}{\nabla_{z} X - \alpha b_{x} \nabla_{y} X} \right|.
\end{equation}
If $\nabla_{z} X < \alpha b_{x} \nabla_{y} X$, then 
the condition of instability is
\begin{equation}
\left| \frac{k_{x}}{k_{z}} \right| > \left| \frac{\alpha b_{z} \nabla_{y} X
+ \nabla_{x} X}{\nabla_{z} X - \alpha b_{x} \nabla_{y} X} \right|.
\end{equation}
Note that the instability occurs even if the magnetic field is weak 
and the magnetization parameter $\alpha$ is small, $\alpha \ll 1$. 
In this case, a surface chemical inhomogeneity can be caused, for 
example, by accretion from the disc and may have a belt-like structure. 
If $\alpha \ll 1$, then from Eq.~(54) we have
\begin{equation}
\left| \frac{k_{x}}{k_{z}} \right| < \left| 
\frac{\nabla_{x} X}{\nabla_{z} X} \right| \sim \frac{H_{X}}{L_{X}},
\end{equation}
and the instability arises only for perturbations with the vertical
wavelength much shorter than the horizontal one. In a moderate 
magnetic field with $L_{X} / H_{X} > \alpha > 1$, inequality (55) yields
\begin{equation}
\left| \frac{k_{x}}{k_{z}} \right| < \left| 
\frac{\alpha b_{z} \nabla_{y} X}{\nabla_{z} X} \right| \sim 
\alpha \frac{H_{X}}{L_{X}},
\end{equation}
and the region of unstable wavevectors is more extended than where
$\alpha \ll 1$. If $\alpha > L_{X} / H_{X}$, then the condition 
of instability is given by Eq.~(56) and reads
\begin{equation}
\left| \frac{k_{x}}{k_{z}} \right| > \left| 
\frac{b_{z}}{b_{x}} \right| \sim 1.
\end{equation}
Note, however, that the condition $\alpha > L_{X}/H_{X}$ requires
a rather strong magnetic field,
\begin{equation}
B > 2 \times 10^{11} Z (1+ x^{2/3}) \frac{L_{X5}}{H_{X3}} \;\;
{\rm G},
\end{equation}
which is impossible in many accreting pulsars ($H_{X3} = H_{X}/
10^{3}$ cm, $L_{X5}= L_{X}/ 10^{5}$ cm).

Condition (50) generalizes the Chandrasekhar criteria of 
stationary convection for the case of a chemically inhomogeneous 
fluid with non-parallel gravity and composition gradient. Since 
$\omega_{A}$ is typically much larger than other characteristic 
frequencies and $\omega_{\kappa} \gg \omega_{\eta}$, it is unlikely 
that inequality (50) is fulfilled in ``standard'' neutron stars because 
these conditions require very low magnetic field and large chemical 
gradient.  

The growth rate of the composition-driven instability ($A_{0} < 0$) is 
always proportional to the gradient of $X$ along the surface.
Taking into account that $\omega_{\kappa} \gg \omega_{\eta}$ and
assuming that the Alfv{\'e}n frequency is much greater than any 
other characteristic frequency, we obtain $|A_{1}|^{2} \gg |A_{0}|$. 
Then, roots 5 and 6 are approximately given by
\begin{equation}
\gamma_{5}^{(1)} \approx - \frac{A_{0}}{A_{1}} \;, \;\;\;
\gamma_{6}^{(1)} \approx - A_{1} \;.
\end{equation} 
As it was mentioned, the condition $A_{1} < 0$ is most likely not
fulfilled in the ocean, which means mode 6 is usually stable. 
Mode 5, however, can be unstable if $A_{0} < 0$. 
Since $\omega_{A}^{2} \gg |\omega_{0}^{2}|$ the growth rate of
this mode is
\begin{eqnarray}
&& \gamma_{5}^{(1)} \approx \frac{\omega_{\eta}}{\omega_{A}^{2}}
( \omega_{X}^{2} + \omega_{XH}^{2}) =
\nonumber \\
&& \frac{4 \pi \rho G \eta_{0} k_{x}}{(\vec{k} \cdot \vec{B})^{2}}
\; [\alpha (\vec{k} \cdot \vec{B}) \nabla_{y} X -
k_{x} \nabla_{z} X + k_{z} \nabla_{x} X].
\end{eqnarray}
We estimate the growth rate for the case when $\vec{k}$ lies in 
the plane perpendicular to the plane $(\vec{G}, \nabla X)$ and, hence, 
$(\nabla X)_{x} =0$. Assuming that the magnetic field is not very 
strong ($L_{X}/H_{X} > \alpha$) and the wavevector satisfies the 
condition of instability (54), we obtain from Eq.~(61)
\begin{equation}
\gamma_{5}^{(1)} \approx \frac{k_{x}}{|\vec{k} \cdot \vec{b}|} \;
\frac{G}{\omega_{pi}} \; |\nabla_{y} X| \sim 10^{-8} \frac{A}{Z} \;
\frac{k_{x}}{k_{z}} \; \frac{g_{14}}{B_{13} L_{X5}} \;\; {\rm s}^{-1}, 
\end{equation}
where $\omega_{pi}=ZeB/Am_{p}c$ is the ion gyrofrequency and $g_{14}=
g/10^{14}$ cm/s$^{2}$. In contrast to the case of oscillatory modes, 
the growth rate of a non-oscillatory mode depends on the direction of 
a wavevector rather than on its value. Instability is  
faster in a neutron star where the surface inhomogeneities of a 
chemical composition are stronger. If the magnetic field is not very 
strong and satisfies the condition $L_{X}/H_{X} > \alpha > 1$
(or $10^{11} Z \Lambda (1+x^{2/3}) L_{X5}/H_{X5} > B >
10^{9} Z \Lambda (1+x^{2/3})$), then the non-oscillatory mode  
grows on a relatively short timescale $\sim 1$ d if $L \sim 100$ m. 
Note that the chemical and thermal gradients are very likely to be 
larger in accreting neutron stars than in isolated ones and one 
could thus expect more efficient hydrodynamic motions in the ocean 
of accreting stars.

\section{Discussion}

It turns out that the ocean of neutron stars is subject to different
instabilities if the temperature or chemical composition varies over 
the surface. The main driving forces of instability are the Hall 
effect and horizontal advection of heat or composition; as a result, 
the growth rate of instability is proportional to the Hall parameter 
$\alpha$ and the component of $\nabla T$ or $\nabla X$ along the 
surface. This point can be clarified by a simple qualitative 
consideration. Consider the case of oscillatory modes in the 
magnetized ocean ($\alpha > 1$) with strong chemical inhomogeneities 
($|\nabla X| \gg \beta |\Delta \nabla T|/T$). Then, the amplitude of 
unstable magnetic perturbations is changed mainly by the Hall effect. 
For example, amplitude $B_{1y}$ increases after $\Delta t$ by
\begin{equation}
\frac{\Delta B_{1y}}{\Delta t} \sim \eta_{\wedge} (\vec{k} \vec{b})
(\vec{k} \times \vec{B}_{1})_{y} \sim \eta_{\wedge} (\vec{k} \vec{b})
\frac{k^{2}}{k_{x}} B_{1z}
\end{equation} 
(using the divergence-free condition (6) for $\vec{B}_{1}$), and small
perturbations are marked by the subscript ``1''. Since oscillatory 
perturbations are approximately Alfvenic in the ocean, the components 
of velocity and magnetic field are related by $B_{1z} \sim B_{1y} 
(v_{1z} / v_{1y})$. For the oscillatory instability, perturbations 
of the composition are relatively small, $X_{1} \propto (\vec{v}_{1} 
\cdot \nabla X) \approx 0$, then $v_{1z} / v_{1y} \sim \nabla X_{y} / 
\nabla Y_{z}$. Substituting these expressions into Eq.~(63), we 
obtain the growth rate (46). It is seen from this simple
consideration that the Hall effect (Eq.~(63)) is one of the
driving forces of instability which makes qualitatively different 
from other instabilities in accreting stars. 

It is worth noting that we neglect rotation in our stability
analysis. Rapid rotation will often change the stability properties 
of accreting neutron stars. The dispersion equation (21) is valid
until angular velocity $\Omega$ is small compared to all
dynamical frequencies. Since $\omega_{A}$ is larger than $\omega_{0}$ 
in many cases of interest, our analysis applies if $\min (\omega_{0},
\omega_{H}) > \Omega$. This condition can be written as $\Omega \ll 
\sqrt{\alpha/(1 + \alpha)} \omega_{0}$ or, using expression (33) 
for $\omega_{0}$, as
\begin{equation}
P > 2 \times 10^{-3} \frac{L_{5}}{H_{3}} \sqrt{ \frac{1+\alpha}{\alpha}}
\;\; {\rm s},
\end{equation}
where $P$ is the rotation period.

The growth time of both the oscillatory and non-oscillatory instabilities 
is relatively short, and instabilities are likely to operate operate in a 
non-linear regime. We can estimate the saturation velocity using the 
mixing-length model (e.g., Schwarzschild 1958) that assumes that the 
turnover time of turbulence generated by instability is on the 
order of the growth time of this instability. Consider, for example,
the saturation regime of oscillatory modes. Then, in a turbulent 
cell with the characteristic vertical lengthscale $\lambda_{z}$, 
the vertical velocity component in saturation can be estimated as 
$v_{Tz} (\lambda_{z}) \sim \lambda_{z} \; {\rm Re} \gamma(\lambda_{z})$. 
Using Eq.~(46), we have
\begin{equation}
v_{Tz} \sim 0.1 B_{13} x^{-1} \epsilon^{-1} \lambda_{z2}^{-1} 
\frac{H_{3}}{L_{5}} \; {\rm cm/s}.
\end{equation} 
Since the instability is anisotropic, the saturated turbulent 
velocity should be anisotropic as well. From Eq.~(4), 
we can estimate the turbulent velocity along the surface as 
$v_{Tx} \sim v_{Tz} (\lambda_{x}/ \lambda_{z})$ where $\lambda_x$
is the characteristic lengthscale of turbulence along the
surface.

Turbulent motions can enhance transport in the ocean both 
vertically and horizontally. For example, turbulence can efficiently
mix the material in the ocean. The coefficient of turbulent 
diffusion can be estimated as a product of the turbulent 
lengthscale and velocity in the corresponding direction. 
Then, we have 
\begin{equation}
\nu_{Tz} \sim v_{Tz} \lambda_{z} \sim 10 B_{13} \epsilon^{-1}
x^{-1} \frac{H_3}{L_5} \; {\rm cm}^{2}/{\rm s}
\end{equation}
for the coefficient of vertical diffusion, and
\begin{equation}
\nu_{Tx} \sim v_{Tx} \lambda_{x} \sim \nu_{Tz} 
\frac{\lambda_{x}^{2}}{\lambda_{z}^{2}}
\sim 10 B_{13} \epsilon^{-1}
x^{-1} \frac{H_3}{L_5} \; \frac{\lambda_{x}^{2}}{\lambda_{z}^{2}}
\; {\rm cm}^{2}/{\rm s}
\end{equation}
for the coefficient of diffusion along the surface. Assuming that 
$\epsilon \approx k_x/k_z \sim H/L$ and $B_{13}=0.1$, $x=1$, we obtain 
$\nu_{Tz} \sim 10^{2}$ cm$^{2}$/s. Such diffusion is sufficient for 
mixing the outer layer with depth $\sim 10$ m on a timescale $\sim$hours. 
The rate of vertical turbulent diffusion should be compared to the 
gravitational sedimentation in the ocean. The coefficient of 
interspecies diffusion (Brown, Bildsten \& Chang 2002) reads in our 
notations as
\begin{equation}
{\cal D} \sim \frac{4 \times 10^{-3}}{A^{0.5} Z^{0.7} Z_{2}^{0.3}}
\; \frac{T_{7}^{1.2}}{x^{0.6}} \; {\rm cm^{2} s^{-1}},
\end{equation}
where $Z_{2}$ is the charge of a trace component. This quantity is 
usually much smaller than $\nu_{Tz}$. As a result, heavy elements
can be transported by turbulence from deep ocean layers to the surface 
and manifest themselves in spectra of both isolated and accreting neutron 
stars. Spectral features should be different in spectra of accreting 
neutron stars with stable and unstable burning. If the accretion rate 
is high and the burning stable, then spectral features corresponding to 
CNO group elements could be detected in such stars. For example, these 
features could be observed in spectra of strongly magnetized accreting 
neutron stars where the burning is stable and a deep ocean of CNO group 
elements is formed (see, e.g., Bildsten 1998). On the other hand, if 
burning is unstable and the star burns H and He directly to iron group 
elements, then mixing caused by instability could increase abundances 
of these elements in the atmosphere and make the corresponding spectral 
features detectable.

Turbulent motions can also lead to an efficient horizontal diffusion 
of the accreted material. The characteristic timescale of turbulent 
diffusion over the neutron star surface can be estimated as $t_{s} 
\sim L^{2}/ \nu_{Tx}$. Substituting expression (67) and assuming 
$\epsilon \sim H/L$, we obtain
\begin{equation}
t_{s} \sim 10^{7} L_{5}^{2} \frac{x}{B_{13}} \left( 
\frac{\lambda_z}{\lambda_x} \right)^{2} \;\; s.
\end{equation} 
If $\lambda_z/\lambda_x = k_x/k_z \sim H/L$ and $B_{13} \sim 0.1$,
$x \sim 1$ then spreading of the accreted material over the surface 
proceeds on a timescale of about a few hours. 

Apart from mixing, turbulence can also enhance the heat transport. 
The turbulent thermal diffusivity is comparable to 
$\nu_{Tz}$, and the ratio of turbulent and electron 
thermal diffusivities is 
\begin{equation}
\frac{\nu_{T}}{\chi_{(0)}} \sim B_{13} T_{7}^{-1} 
(1 + x^{2/3}) x^{-1} \lambda_{\rm z2}^{-1}.
\end{equation}
In general, these quantities can have the same order of magnitude. 
Since $\nabla T$ is subadiabatic, turbulent motions increase the 
difference between the surface and internal temperature. Note that
turbulent heat transport along the surface and vertically should 
be different for the considered instability with more efficient 
diffusion along the surface, which can reduce the surface temperature 
gradient and decrease the contrast between the polar and equatorial 
temperature.

The non-oscillatory instability, despite being slower than the oscillatory
one, can also manifest itself in accreting neutron stars. A sufficiently 
strong magnetic field can efficiently funnel the accreted material 
onto the magnetic pole. The characteristic radius of such accretion 
spots in the polar region is likely to be substantially
smaller than the stellar radius, $L \sim 10^{4}-10^{5}$ cm. In a 
moderate magnetic field satisfying the inequality $L_{X}/H_{X} > 
\alpha > 1$, the growth rate can be estimated if we suppose in Eq.~(62) 
that $k_x/k_z \sim \alpha H_{X}/L_{X}$ (see Eq.~(57)). Then,
\begin{equation}
\gamma_{5}^{(1)} \sim 10^{-6} (1+x^{2/3})^{-1} \frac{A}{Z^{2} \Lambda}
\cdot \frac{g_{14} H_{X3}}{L_{X5}^{-2}} 
\;\;\;\; {\rm s}^{-1}.
\end{equation}
If $Z$ is not very large and the radius of accretion spots on the 
surface is $\sim 100$ m, then the growth time of instability can reach 
$10^{4}-10^{5}$ s. In the non-linear regime, the turbulent velocity 
induced by this instability is rather small, $\sim 0.1-0.01 
L_{X5}^{-1}$ cm/s, if $x \sim g_{14} \sim H_{X3} \sim 1$. Slow turbulent
motions induced by the non-oscillatory instability in the spot can
produce slow variations in the X-ray luminocity of accreting pulsars 
on the same timescale, which should be longer for accreting 
neutron stars with a weaker magnetic field. 

Note that the criteria and growth rates of instability have been 
derived under the assumption $c_{s} > c_{A}$ or $p > B^{2} / 4 \pi$. 
Neglecting the quantum effects, we have $p \approx 9.7 \times 10^{22}
x^{5/3}$ dyne/cm$^{2}$ in the region where the electron gas is 
non-relativistic and $x \leq 1$. Then, our results applies if
\begin{equation}
B < B_{cr}(x) \approx 1.1 \times 10^{12} x^{5/6} \; {\rm G}.
\end{equation}
Stability properties of the layers where $c_{A} > c_{s}$ (or $B> 
B_{cr}$) are different from those considered in our study. 
Generally, perturbations of $T$ and $X$ can be suppressed for modes
with $\gamma \sim \omega_{A} > \omega_{s}$. As a result, a strong
magnetic field in the region where $c_{A} > c_{s}$ can reduce the 
growth rate of instabilities by a factor $(c_{A}/c_{s})^{2}$. 
Therefore, the turbulent transport on a short time-scale can be 
suppressed in these layers, and instability is less efficient in 
the surface region where $c_{A} > c_{s}$ or    
\begin{equation}
\rho < \rho_{cr} \approx 2.8 \times 10^{7} B_{13}^{6/5} {\rm g/cm^{3}}.
\end{equation}
A difference in turbulent transport between the regions with $\rho >
\rho_{cr}$ and $\rho < \rho_{cr}$ could produce a qualitative difference 
in the burst activity of strongly and weakly magnetized accreting neutron 
stars. The thermonuclear X-ray burst is triggered when H/He burning 
becomes unstable at the base of the accumulated layer, at a density 
$\sim 10^{5}-10^{6}$ g/cm$^{3}$ (see, e.g., Bildsten 1998, Cumming \& 
Bildsten 2000). However, strongly magnetized accreting neutron stars 
are not observed among X-ray bursters, despite accretion at rates for 
which the burning of H/He should be unstable. This is usually considered 
as evidence that material is not spreading over the surface on short 
timescales, so the local accretion rate is high enough to stabilize the 
burning (Joss \& Li 1980). In this model, the burning can be stabilized 
if spreading is suppressed only in the surface layer above the region  
with $\rho \sim 10^{5}- 10^{6}$ g/cm$^{3}$ (or $x < 0.1-1$) where the 
H/He burning occurs. One can estimate from Eq.~(72) that instability 
and mixing become less efficient in the layers with $\rho < 10^{5}-
10^{6}$ g/cm$^{3}$ if 
\begin{equation}
B > B_{cr}(x \sim 0.1-1) \sim 10^{11}-10^{12} \;\; {\rm G}.
\end{equation}   
Therefore, one can expect that the accreted material is spreading over
the surface layer with $\rho < 10^{5}-10^{6}$ g/cm$^{3}$ on a short
timescale only in weakly magnetized neutron stars with $B<10^{11}-
10^{12}$ G, and the H/He burning in such stars should be unstable. On 
the contrary, in strongly magnetized neutron stars with $B>10^{11}-
10^{12}$ G, the spreading over the surface layer requires a longer
time, and these neutron stars will not manifest themselves as X-ray
bursters. The difference between strongly and weakly magnetized
accreting neutron stars will be considered in more detail in a
forthcoming paper.   

\vspace{0.2cm}
\noindent{\it Acknowledgements.}
This work was supported by the Spanish Ministerio de Ciencia 
y Tecnologia (grant AYA 2001-3490-C02).

\end{document}